\documentclass[twocolumn,usenatbib,usegraphicx]{mn2e}

\def\mnras{MNRAS}

\def\aj{AJ}
\def\apj{ApJ}
\def\apjl{ApJL}
\def\apjs{ApJS}
\def\aap{A\&A}

\usepackage{epsfig}
\usepackage{amssymb}
\newcommand{\mpch}{\ensuremath{{\rm Mpc\,}h^{-1}}}

\begin{document}

\title{The size of the longest filament in the Luminous Red Galaxy distribution}

\author[B. Pandey, G. Kulkarni, S. Bharadwaj \& T.  Souradeep]
       {Biswajit Pandey$^1$\thanks{Email: biswap@visva-bharati.ac.in},
         Gauri Kulkarni$^2$\thanks{Email: gaurik@iucaa.ernet.in},
         Somnath Bharadwaj$^3$\thanks{Email: somnathb@iitkgp.ac.in} \&
         Tarun Souradeep$^2$\thanks{Email: tarun@iucaa.ernet.in}
         \\${}^1$ Department of Physics, Visva-Bharati University,
         Santiniketan, Birbhum, 731235, India\\ ${}^2$
         Inter-University Centre for Astronomy and Astrophysics, Pune
         411007, India \\${}^3$ Department of Physics and Meteorology
         and Centre for Theoretical Studies, IIT Kharagpur, 721302,
         India}
\maketitle

\begin{abstract}
Filaments are one of the most prominent features visible in the galaxy
distribution. Considering the Luminous Red Galaxies (LRGs)  in the 
Sloan Digital Sky Survey Data Release Seven (SDSS DR7), we have
analyzed the  filamentarity in $11$ nearly two dimensional (2D)
sections through a volume limited subsample of this data. The
galaxy distribution, we find,  has excess  filamentarity in
comparison to a random distribution of points.  We use 
a  statistical technique ``Shuffle'' to determine  $L_{\rm MAX}$, the
largest length-scale at which we have statistically significant
filaments. We find that $L_{\rm MAX}$ varies in the range $100-130 
\, h^{-1} {\rm Mpc}$ across the $11$ slices, with a mean value 
 $L_{\rm MAX}=110 \pm 12 \, h^{-1}{\rm Mpc}$. Longer filaments,
though possibly present in our data, are not statistically significant
and are the outcome of chance alignments. 
\end{abstract}


\section{Introduction}

One of the main goals of  redshift surveys is to study the
galaxy distribution.  Various redshift surveys
(e.g.  CfA, \citealt{gel}; LCRS, \citealt{shect}; 2dFGRS,
\citealt{colles} and SDSS, \citealt{stout}) all show that galaxies are
distributed in an interconnected network of clusters, sheets and
filaments encircling nearly empty voids. This complex network is often
referred to as the ``Cosmic Web''. Quantifying the cosmic web and
understanding it's origin is a challenging problem in cosmology.

The analysis of filamentary patterns in the galaxy distribution has a
long history dating back to a few papers in the late-seventies and
mid-eighties by \citet{joe}, \citet{einas4}, \citet{zel},
\citet{shand1} and \citet{einas1}. Filaments are the most striking
visible patterns seen in the galaxy distribution (e.g. \citealt{gel},
\citealt{shect}, \citealt{shand2}, \citealt{bharad1}, \citealt{mul},
\citealt{basil}, \citealt{doro2}, \citealt{pimb}, \citealt{pandey}).

Are the filaments statistically significant? This is a question that
naturally arises when  we embark on studying the filaments in the
galaxy distribution.  The possibility that the observed
filaments are not genuine feature of the galaxy distribution and 
could arise out of chance alignments requires us to establish the
statistical significance of the filaments.

The SDSS \citep{york} is currently the largest galaxy redshift
survey. \citet{pandey} (hereafter Paper I) have analyzed the
filamentarity in the equatorial strips of this survey. These strips
are nearly two dimensional (2D). They have projected the data onto a 
plane and analyzed the resulting 2D galaxy distribution. They find
evidence for connectivity and filamentarity in excess of that of a
random point distribution, indicating the existence of an
interconnected network of filaments. The filaments are
statistically significant upto a length scales of $80 \, h^{-1} {\rm
  Mpc}$ and not beyond \citep{pandey}. All the structures spanning  
length-scales greater than this length scale are the result of chance
alignments. These results are consistent with the earlier findings
from Las Campanas Redshift Survey (LCRS) where the filamentarity was
found to be statistically significant on scales up to $70-80 \, h^{-1}
\, {\rm Mpc}$ in the $-3^{\circ}$ slice and $50-70 \, h^{-1} \, {\rm
  Mpc}$ in the other 5 slices \citep{bharad2}. The average
filamentarity of the galaxy distribution was  shown to depend on
various physical galaxy properties  such as the luminosity, colour,
morphology and star formation rate \citep{pandey1, pandey3}. Recently
\citet{pandey4} has studied if the statistically significant length scale
of filaments depends on different galaxy properties and finds that it
does not depend on the galaxy luminosity, colour and
morphology. Analysis with mock galaxy samples from N-body simulations  
indicates that the length scale upto which filaments are statistically
significant is also nearly independent of bias and weakly depends on
the number density and size of the galaxy samples.

The measurement of the length scale upto which the filaments are
statistically significant could be limited by the size of the survey,
and it is possible  that this has been underestimated  in earlier
studies with the LCRS and the SDSS MAIN galaxy sample. This issue can
be addressed  repeating the analysis with a bigger galaxy survey. 
Given a magnitude limit, Luminous Red Galaxies  (LRGs) can be
observed to greater distances as compared to normal $L_{\star}$
galaxies. Further their stable colors make them 
relatively easy to pick out from the rest of the galaxies using the
SDSS multi-band photometry. The LRG sample extends to a much deeper
region of the Universe as compared to the SDSS MAIN galaxy sample. The
very large region covered by the SDSS LRG sample provides us the ideal
opportunity to investigate the size of the longest filaments present
in the Universe.

In this work we measure the largest lengthscale at which we have 
statistically significant filaments in the distribution of 
LRGs in the SDSS DR7.  We discuss the data and method of analysis
in Section 2,  and the results and conclusions are presented in
Section 3. 

Throughout our work, we have used the flat $\Lambda$CDM cosmology with
$\Omega_m = 0.3, \; \Omega_{\Lambda} = 0.7 \mbox{ and } h = 1$.

\section{Data and Method of Analysis}

\subsection{SDSS data}

The Sloan Digital Sky Survey (SDSS) \citep{york} is a wide-field
imaging and spectroscopic survey of the sky using a dedicated 2.5 m
telescope \citep{gunn} with $3^{\circ}$ field of view at Apache Point
Observatory in southern New Mexico. The SDSS has imaged the sky in 5
passbands, $u,\; g,\; r,\; i$ and $z$ covering $10^4$ square degrees
and has so far made 7 data releases to the community. The SDSS galaxy
sample can be roughly divided into two (i) the MAIN galaxy sample and
(ii) the Luminous Red Galaxy (LRG) sample. For our work, we have used the
spectroscopic sample of LRGs derived from the Data Release 7 of
SDSS-II \citep{abaz}.

SDSS targets those galaxies for spectroscopy which have $r$ magnitude
brighter than 17.77 ($r < 17.77$). To select LRGs, additional galaxies
are targeted using color-magnitude cuts in $g,\; r$ and $i$ which
extends the magnitude limit for LRGs to $r < 19.5$. The prominent
feature for an early-type galaxy is the $4000 \, A^{\circ}$ break in its
SED. For $z \lesssim 0.4$ this feature lies in the  $g$ passband  while it
shifts to the $r$ band for higher redshifts. Hence selection of LRGs
involves different selection criteria below and above $z \lesssim
0.4$, the details of which are described in \citet{Eisenstein2001}
(hereafter E01).

The criteria for lower redshifts are collectively called Cut-I (eq. (4-9)
in E01) while those for the higher redshifts are called Cut-II (eq. 
(9-13) in E01), with Cut-I accounting for $\approx 80\%$ of the
targeted LRGs. A galaxy that passes either of these cuts is flagged
for spectroscopy by the SDSS pipeline as TARGET\_GALAXY\_RED while a
galaxy that passes only Cut II is flagged as
TARGET\_GALAXY\_RED\_II. Hence, while selecting the 
LRGs we require that the TARGET\_GALAXY\_RED and TARGET\_GALAXY\_RED\_II
flags both  be set.

\begin{figure}
\rotatebox{-90}{\scalebox{.4}{\includegraphics{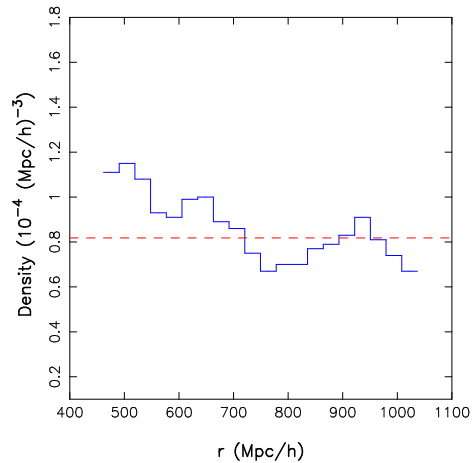}}} 
\caption{The density of LRGs as a function of radial
  distance $r$. The density has been  computed using  shells of
  uniform   thickness   $28.75 h^{-1} {\rm Mpc}$ in the radial
  direction. The dashed line shows the mean  density of our  LRG
  sample.} 
  \label{fig:1}
\end{figure}

We obtain the $g$-band absolute magnitude from the $r$ band apparent
magnitude accounting for the k-correction and passive evolution. We
use the prescription for $K+e$ correction from the Table 1 in E01
(non-star forming model). The cuts defined above are designed to
produce an approximately volume limited sample of LRGs upto $z \approx
0.4$. The comoving number density of LRGs falls sharply beyond $z
\approx 0.4$. It is obviously necessary to include LRGs that come from
the MAIN sample for $z \lesssim 0.3$. But E01 issue a strong advisory
against selecting LRGs with $z \lesssim 0.15$ for a volume limited
sample as the luminosity threshold is not preserved for low
redshifts. We have restricted our LRG sample to the redshift range
$0.16 < z < 0.38$ and $g$-band absolute magnitude range $-23<M_g<-21$.
Figure \ref{fig:1} shows the LRG number density as a function of the
radial distance $r$ for the $r$ range that we have used in our
analysis. We see that the LRG density has a maximum variation of $\sim
30 \%$ around the mean density.  An earlier study \citep{pandey1}
indicates that a density variation of this order is not expected to
produce a statistically significant effect on the average
filamentarity. The earlier study also indicates that the length scale
upto which the filaments are statistically significant has a very weak
dependence on the density.

\begin{figure}
\rotatebox{0}{\scalebox{.4}{\includegraphics{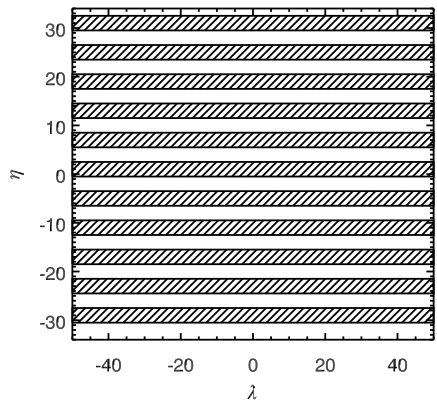}}} 
\caption{The geometry of the 11 LRG strips extracted from SDSS
  DR7 plotted in survey coordinates.}
  \label{fig:2}
\end{figure}

Figure \ref{fig:2} shows the geometry of the sky coverage of the
region that we have analyzed  in survey coordinates $\lambda$ and
$\eta$ \citep{stout}. For our analysis we have 
extracted $11$ strips that are described in the Table~1. Each strip
spans $100^\circ$ in $\lambda$ ($-50<\lambda<50$) and $3^\circ$ in
$\eta$ ($\eta$ ranges are shown in Table~1), and radially extends from
$462h^{-1} {\rm Mpc}$ to $1037  h^{-1} {\rm Mpc}$. Adjacent
strips are separated by $3^\circ$. The physical thickness of each
strip increases with radial distance. We have extracted strips of
uniform thickness   $24 h^{-1} {\rm Mpc}$ which
corresponds to $3^{\circ}$ at a radial distance of 
$462h^{-1} {\rm Mpc}$. The volume outside these uniform thickness
strips was discarded from our analysis.

\begin{figure*}
  \rotatebox{0}{\scalebox{1.}{\includegraphics{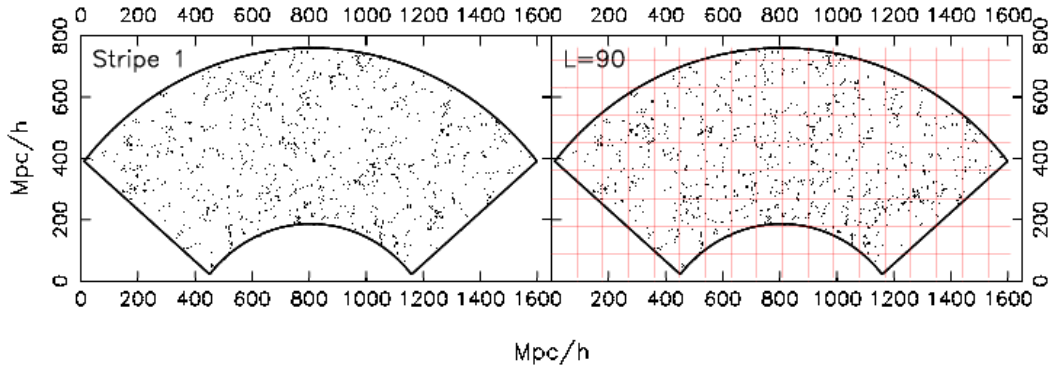}}}
  \caption{ The left panel shows the projected 2D galaxy distribution
    in one of the LRG strips. The right panel shows a shuffled
    realization generated from the same data using $L=90 h^{-1} {\rm
      Mpc}$. The corresponding $90 h^{-1}{\rm Mpc} \, \times \, 90
    h^{-1} {\rm Mpc}$ grid is overlaid. The blocks defined by this
    grid were randomly interchanged with a rotation. Only blocks that
    are entirely within the survey area are used in this process.}
\label{fig:4}
\end{figure*}

\begin{table}{}
\caption{Details of the LRG samples extracted from SDSS DR7}
\begin{tabular}{|c|c|c|c|c|}
\hline
Strip name &$\eta_{min}$ & $\eta_{max}$ & $n$ \\
\hline
S1 & -33.5 &  -30.5 &   1211\\

S2 & -27.5 &  -24.5 &   1332\\

S3 & -21.5 &  -18.5 &   1477\\

S4 & -15.5 &  -12.5 &   1513\\

S5 & -9.50 &  -6.50 &   1590\\

S6 & -3.50 & -0.500 &   1568\\

S7 &  2.50 &   5.50 &   1677\\

S8 &  8.50 &   11.5 &   1601\\ 

S9 &  14.5 &   17.5 &   1422\\

S10 &  20.5 &   23.5 &  1169\\

S11 &  26.5 &   29.5 &  1356\\

\hline
\end{tabular}
\end{table}

\subsection{Method of Analysis}
The strips that we have analyzed extend $575 \, h^{-1} {\rm Mpc}$ in
the radial direction and $\sim 800 \, h^{-1} {\rm Mpc}$ (or more) in
the transverse direction, while the thickness is only $24 \, h^{-1}
{\rm Mpc}$. These strips can be treated as nearly two dimensional
which makes the analysis relatively simpler.  The strips were all
collapsed along the thickness (the smallest dimension) to produce a 2D
galaxy distributions (Figure \ref{fig:4}).  We use the 2D
``Shapefinder'' statistic \citep{bharad1} to quantify the
filamentarity of the patterns in the resulting galaxy distribution.  A
detailed discussion is presented in \citet{pandey1}, and we present
only the salient features here. The reader is referred to
\citet{sahni} for a discussion of Shapefinders in three dimensions
(3D).

 The galaxy distribution was embedded in a 2D mesh $760  \, h^{-1} {\rm
   Mpc} \times 1608 \, h^{-1} {\rm Mpc}$ with cell size  $2 \, h^{-1}
 {\rm Mpc} \times 2 \,h^{-1}  {\rm Mpc}$. The entire galaxy
 distribution was represented as a set of 1s and 0s on the mesh. 
Cells containing  a galaxy were  assigned a value $1$, and the empty
cells were assigned a value $0$.  Note the  difference in  grid
size from the \citet{pandey} where a  $1\,\mpch \times 1\, \mpch$
cell was used. This takes into account the fact that our LRG strips 
have  projected galaxy number density  $\simeq 1.9 \times 10^{-3}
(\mpch)^{-2}$which is considerably smaller than  $\simeq 9 \times
10^{-3} (\mpch)^{-2}$ in  the samples analyzed in Paper I.

We use the `Friends-of-Friend'  (FOF) algorithm to identify connected
 regions of filled  cells which we refer to as clusters. The
 filamentarity of each cluster is quantified  
 using the Shapefinder ${\cal F}$ which defined as
\begin{equation}
{\cal F} = \frac{(P^2 - 16 S)}{(P-4 l)^2}
\end{equation}
 where $P$ and $S$ are respectively the perimeter and the area of the
 cluster, and $l$ is the grid spacing. The  Shapefinder ${\cal F}$
 has  values $0$ and $1$ for a square and filament respectively, and
 it assumes intermediate values as  a  square is deformed to a
 filament. We use the average filamentarity  
\begin{equation} 
F_2 = {\sum_{i} {\cal S}_i^2 {\cal F}_i\over\sum_{i}{\cal S}_i^2} \,. 
\end{equation}
to asses the overall filamentarity of  the clusters in the galaxy
distribution. 

\begin{figure}
  \rotatebox{-90}{\scalebox{0.4}{\includegraphics{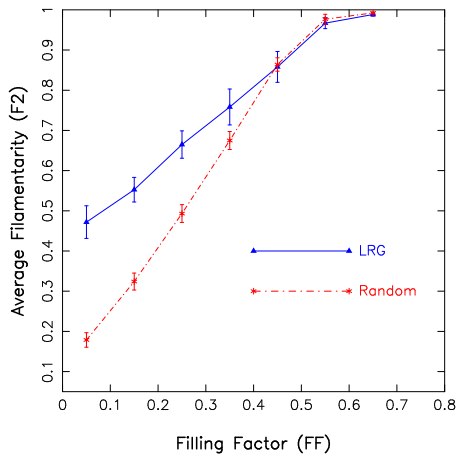}}}
  \caption{The Average Filamentarity ($F_2$) as a function of the 
Filling Factor ($FF$).  The mean and $1-\sigma$ errors from $11$
strips are shown for both the LRG strips and the corresponding random
realisations. }
\label{fig:5}
\end{figure}

The distribution of 1s corresponding to the galaxies is sparse. Only
$\lesssim 1 \%$ of the cells contain galaxies and there are very few
filled cells which are interconnected.  As a consequence FOF fails to
identify the large coherent structures which correspond to filaments
in the galaxy distribution. We overcome this by successively
coarse-graining the galaxy distribution. This is achieved by gradually
making the filled cells fatter. In each iteration of coarse-graining
all the empty cells adjacent to a filled cell (i.e. cells at the 4
sides and 4 corners of a filled cell) are assigned a value $1$ . This
causes clusters to grow, first because of the growth of individual
filled cells, and then by the merger of adjacent clusters as they
overlap. Coherent structures extending across progressively larger
length-scales are identified in consecutive iterations of
coarse-graining.  So as not to restrict our analysis to an arbitrarily
chosen level of coarse-graining, we study the average filamentarity
after each iteration of coarse-graining. The filling factor $FF$
quantifies the fraction of cells that are filled and its value
increases from $\sim 0.01$ and approaches $1$ as the coarse-graining
proceeds. We study the average filamentarity $F_2$ as a function of
the filling factor $FF$ (Figure \ref{fig:5}) as a quantitative measure
of the filamentarity at different levels of coarse-graining. The
values of $FF$ corresponding to a particular level of coarse-graining
shows a slight variation from strip to strip.  In order to combine and
compare the results from different strips, for each strip we have
interpolated $F_2$ to $7$ values of $FF$ at a uniform spacing of $0.1$
over the interval $0.05$ to $0.65$. Coarse-graining beyond $FF \sim
0.65$ washes away the filaments and hence we do not include this range
for our analysis.  For comparison we also consider a random reference
sample generated by randomly distributing points over a 2D region with
exactly the same geometry and number density as the projected LRG
samples.  We find that the filamentarity of the LRG data is in excess
of that of the random samples (Figure \ref{fig:5}). While this is
reassuring that we are studying a genuine signal, it should be
interpreted with some caution. It is possible that the enhanced
filamentarity observed in the actual data would also arise if the
observed two-point clustering were incorporated in the random samples
even in the absence of higher order clustering. In this paper we use a
statistical technique called Shuffle, which does not rely on
externally generated random samples, to establish the length-scale
upto which the observed filamentarity is statistically significant.
Shuffle was first introduced and applied by \citet{bhav}. Subsequent
work \citep{bharad2,pandey,pandey4} has extended this and applied it
to the LCRS, the SDSS Main galaxy sample and N-body simulations.

A grid with squares blocks of side $L$ is superposed on the original
data slice (Figure \ref{fig:4}). The blocks which lie entirely within
the survey area are then randomly interchanged, with rotation,
repeatedly to form a new shuffled data. The shuffling process
eliminates coherent features in the original data on length-scales
larger than $L$, keeping structures at length-scales below $L$ intact.
All the structures spanning length-scales greater than $L$ that exist
in the shuffled slices are the result of chance alignments. At a fixed
value of $L$, the average filamentarity in the original sample will be
larger than in the shuffled data only if the actual data has more
filaments spanning length-scales larger than $L$, than that expected
from chance alignments. The largest value of $L$, $L_{\rm MAX}$, for
which the average filamentarity of the shuffled slices is less than
the average filamentarity of the actual data gives us the largest
length-scale at which the filamentarity is statistically
significant. Filaments spanning length-scales larger than $L_{\rm
  MAX}$ arise purely from chance alignments.

 For each value of $L$ we have generated $24$ different realization of the
 shuffled slices.  To ensure that the edges of the blocks which are
 shuffled around do not cut the actual filamentary pattern at exactly
 the same place in all the realizations of the shuffled data, we have 
 randomly shifted the origin of the grid used to define the blocks.
 The values of FF and $F_2$ in the 24 realizations differ from one
 another and from the actual data at the same stage of
 coarse-graining. So as to be able to quantitatively compare the
 shuffled realizations with the actual data, results from the shuffled
 data were interpolated in the same way as actual data.  The mean 
 $\bar{F_2}[{\rm Shuffled}]$ and the variance $(\Delta F_2[{\rm
     Shuffled}])^2$ of the average filamentarity was determined for
 the shuffled data at each value of FF using the $24$ realizations.  The
 difference between the filamentarity of the  actual data and its
 shuffled counterparts was quantified using the reduced $\chi^2$ per
 degree of  freedom
\begin{equation}
\frac{\chi^2}{\nu}=\frac{1}{N_p} \sum_{a=1}^{N_p}
 \frac{(F_2[{\rm Actual}]-\bar{F_2}[{\rm Shuffled}]))_a^2 }
{(\Delta  F_2[{\rm Shuffled}])_a^2} 
\end{equation}
where the sum is over different values of the filling factor FF. 

\begin{center}
\begin{figure}
 \rotatebox{-90}{\scalebox{0.4}{\includegraphics{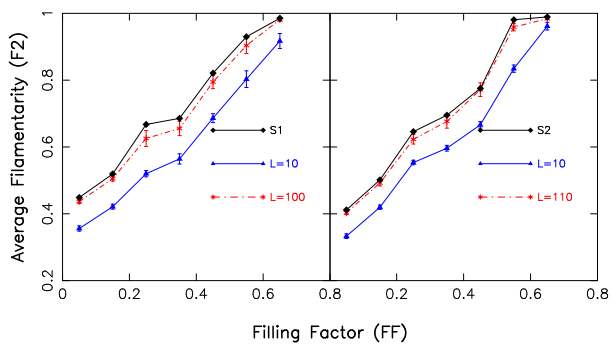}}}
 \caption{The Average Filamentarity ($F_2$) as a function of the 
 Filling Factor ($FF$) for strips S1 (left) and S2 (right) of the LRG
 data.  The effect of shuffling with $L=10 h^{-1} \, {\rm Mpc}$ is
 shown for both strips. The other value of $L$ corresponds to $L_{\rm
 MAX}$ which is different for the two strips. Notice that at $L_{\rm
 MAX}$ the actual  data points lie just outside the $1-\sigma$
 error-bars of  the shuffled data. For the shuffled data we have shown
 the mean and $1-\sigma$ errors from $24$ independent realizations of
 the shuffling process.} 
  \label{fig:6}
\end{figure}
\end{center}

\begin{center}
\begin{figure}
 \rotatebox{-90}{\scalebox{0.3}{\includegraphics{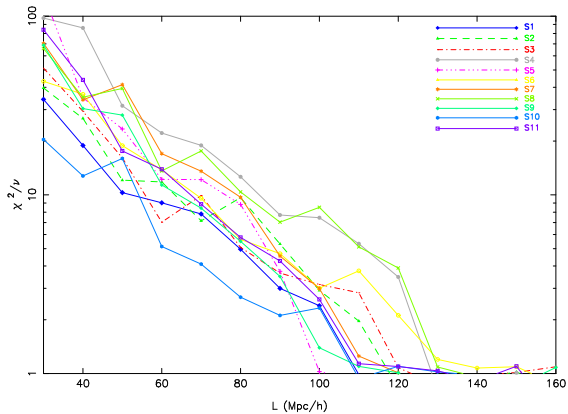}}}
 \caption{ This shows $\chi^2/\nu$ as a function of $L$  for
   all the $11$  LRG strips that we have analyzed.}
 \label{fig:7}
\end{figure}
\end{center}

\section{Results and Conclusions}
Figure \ref{fig:6} shows the effect of shuffling the LRG data for $2$
of the $11$ strips. We see that shuffling with $L=10 \, h^{-1} {\rm
  Mpc}$ causes $F_2$ to drop considerably relative to the unshuffled
data. It is thus established that the filaments are statistically
significant at this length-scale. We have varied $L$ in steps of $10
\, h^{-1} {\rm Mpc}$ form $L= 10 \, h^{-1} {\rm Mpc}$ to $160 \,
h^{-1} {\rm Mpc}$.  We find that the difference between the actual and
the shuffled data is reduced ({\it ie.}  $F_2$ increases) as $L$ is
increased. The filamentarity in the actual data and its shuffled
counterparts are consistent for $L=160 \, h^{-1} {\rm Mpc}$ indicating
that at this length-scale the filaments are not statistically
significant but are the outcome of chance alignments. All the $11$ LRG
strips analyzed here show a similar behaviour.

We consider $\chi^2/\nu$ to quantify the difference between the
filamentarity of the actual data and its shuffled counterparts.  This
difference is not considered to be statistically significant if
$\chi^2/\nu \sim 1$.  For all the $11$ strips, Figure \ref{fig:7}
shows $\chi^2/\nu$ for different values of $L$.  The value of
$\chi^2/\nu$ falls with increasing $L$.  We define $L_{\rm MAX}$ as
the largest $L$ value for which there is a statistically significant
difference caused by shuffling.  Increasing $L$ beyond $L_{\rm MAX}$
causes $\chi^2/\nu$ to fall such that $\chi^2/\nu \sim1$.  Considering
the particular strip shown in the left panel of Figure \ref{fig:6}, we
see that $L_{MAX}= 100 \, h^{-1} {\rm Mpc}$ i.e. there is a
statistically significant difference between the average filamentarity
of the actual data and its shuffled counterparts for $L \le L_{\rm
  MAX}= 100 \, h^{-1} {\rm Mpc}$ but not beyond.  The value of
$L_{MAX}$ varies in the range $100-130$ across the $11$ strips that we
have analyzed.  We have averaged the values of $L_{MAX}$ measured in
the $11$ different strips to find that $L_{\rm MAX}=110 \pm 12 \,
h^{-1} {\rm Mpc}$.  This is the largest length-scale at which we have
statistically significant filaments in our LRG strips. Filaments
longer than this, though possibly present in the data, are the outcome
of chance alignments.

The value of $L_{\rm MAX}$ estimated using the LRGs  is somewhat
larger than the values in the range  $70-80 \, h^{-1} {\rm Mpc}$
obtained in earlier studies using the  LCRS \citep{bharad2} and the
SDSS MAIN galaxy sample  \citep{pandey}. This is possibly a
consequence of the fact that LRG strips cover a much larger area as
compared to the ones analyzed earlier. For example, the SDSS 
strips analyzed in \citet{pandey} have an extent of $336$ and $373 \,
h^{-1} {\rm Mpc}$ in the radial and transverse directions respectively,
in comparison to $575$ and $800  h^{-1} {\rm Mpc}$  for the strips
analyzed here. The larger area ensures a better  mixing of the blocks
in the shuffling process. This is particularly important at large $L$
where we had very few blocks in the smaller strips that had been
analyzed earlier.  The larger area also ensures that the  present
estimate of $L_{\rm MAX}$ is less likely to be influenced by local
effects, and hence is more representative of the global value. 

Finally we note that tests with controlled mock samples
\citep{pandey4} show that the 2D analysis adopted here tends to
shorten the length of 3D filaments. Further, 3D sheets will appear as
filaments in 2D. It is thus necessary to be cautious in interpreting
the consequence of our finding in terms of the full 3D galaxy
distribution. It is  reasonable to interpret $L_{\rm MAX}=110 \pm
12 \, h^{-1} {\rm Mpc}$ as an order of magnitude estimate of the
largest length-scale at which we have statistically significant
structures (sheets and  filaments) in the 3D LRG  distribution. 

\section{Acknowledgment}

Computations were carried out at the IUCAA HPC facility. GK
acknowledges support through a postdoctoral position under the
DST-Swarnajayanti fellowship grant of TS. GK acknowledges James Annis
for his help in understanding the LRG data.

The SDSS DR7 data was downloaded from the SDSS skyserver
http://cas.sdss.org/dr7/en/.

    Funding for the creation and distribution of the SDSS Archive has been 
provided by the Alfred P. Sloan Foundation, the Participating 
Institutions, the National Aeronautics and Space Administration, the 
National Science Foundation, the U.S. Department of Energy, the Japanese 
Monbukagakusho, and the Max Planck Society. The SDSS Web site is 
http://www.sdss.org/.

    The SDSS is managed by the Astrophysical Research Consortium (ARC) for 
the Participating Institutions. The Participating Institutions are The 
University of Chicago, Fermilab, the Institute for Advanced Study, the 
Japan Participation Group, The Johns Hopkins University, the Korean 
Scientist Group, Los Alamos National Laboratory, the Max-Planck-Institute 
for Astronomy (MPIA), the Max-Planck-Institute for Astrophysics (MPA), New 
Mexico State University, University of Pittsburgh, Princeton University, 
the United States Naval Observatory, and the University of Washington.





\end{document}